\documentclass[proceedings,nohyper]{JHEPN}
\usepackage{epsfig}
\newcommand{\bee}{\begin{equation}}
\newcommand{\ene}{\end{equation}}
\newcommand{\bea}{\begin{array}}
\newcommand{\ena}{\end{array}}
\newcommand{\beqa}{\begin{eqnarray}}
\newcommand{\enqa}{\end{eqnarray}}
\newcommand{\bean}{\begin{eqnarray*}}
\newcommand{\eean}{\end{eqnarray*}}

\newcommand{\del}{\partial}

\newcommand{\jou}[4]{{\rm #1 }{\bf #2} (#3) #4}
\renewcommand{\jou}[4]{{\rm #1 }{\bf #2} (#3) #4}
\def\ru1{\rule[-0.4truecm]{0mm}{1truecm}}

\conference{\hspace{-3truemm} A Semianalytical Method to Solve Altarelli-Parisi Evolution Equations}

\title{A SEMIANALYTICAL METHOD TO SOLVE ALTARELLI-PARISI EVOLUTION
EQUATIONS\thanks{Based on talk given by the first author at the 6th Hellenic
School and Workshop on Elementary Particle Physics, Corfu, Greece, September 1998.
Published in {\it JHEP} Conf.Proc. corfu98/023}}

\author{P. Santorelli and E. Scrimieri\\
$^{a}$Dipartimento di Scienze Fisiche, Universit\`a "Federico II"
di Napoli\\ Mostra d'Oltremare, Pad. 20, I-80125 Napoli, and INFN,
Sezione di Napoli, Italy\\ $^{b}$Dipartimento di Fisica,
Universit\`a di Bari,\\ Via G. Amendola, 173, I-70126 Bari, and
INFN, Sezione di Bari, Italy.}

\abstract{We discuss a new method to solve in a semianalytical way
the Dokshitzer-Gribov-Lipatov-Altarelli-Parisi evolution equations
at NLO order in the $x$-space. The method allows to construct an
evolution operator expressed in form of a rapidly convergent
series of matrices, depending only on the splitting functions.
This operator, acting on a generic initial distribution, provides
a very accurate solution in a short computer time (only a few
hundredth of second). As an example, we apply the method, useful
to solve a wide class of systems of integrodifferential equations,
to the polarized parton distributions.\\ \hfill
\\
Bari-TH/99-358\hspace{0.5truecm} DSF 31/99}

\begin{document}

\section{Introduction}
\label{s:intro}

The scaling violation of nucleon structure functions is described in
terms of Dokshitzer-Gribov-Lipatov-Altarelli-Parisi (DGLAP) evolution
equations \cite{DGLAP}. The DGLAP integrodifferential equations describe
the $Q^2$ dependence of the structure functions, which are related, {\it
via} the operator product expansion,  to the parton distributions for
which the DGLAP equations are usually written down. In this framework,
the analysis of the experimental data, is performed fixing at some
$Q_0^2$ the structure functions by assuming the parton distributions and
computing the convolution with the coefficient functions, which can
be evaluated in perturbation theory. The comparison with experimental
data, which are distributed at different values of $Q^2$, goes through
the solution of DGLAP equations for the parton distributions; thus a
reliable and fast algorithm to solve these equations is welcome.

In literature there are essentially three different approaches to
solve the DGLAP equations. The first one is based on the Laguerre
polynomials expansion \cite{FurmPetr}. This technique is quite
accurate up to $x$-values not smaller than $\bar x \approx
10^{-3}$; on the contrary, below $\bar x$ the convergence of the
expansion slows down \cite{kumLon,CorianoSavkli}. Given that
experimental data are already available down to about $\bar x$,
for the polarized case, and down to $10^{-5}$, for the unpolarized
case, this method results no longer practical.\\ An alternative
approach takes advantage of the fact that the moments of the
convolutions appearing in the equations factorize in such a way
that the analytical solution, in the momentum space, can be
written down \cite{Glucketal}. However, also in the most favorable
case in which the analytical expressions of the moments of the
initial conditions are known, the numerical Mellin inversion is
relatively CPU time consuming (see \cite{SMCcomparison}).
Moreover, as discussed in \cite{Forte}, since $x$ variable is
related to the invariant energy $W^2$ of the virtual photon-hadron
scattering process by $W^2 =(1-x)/x$, $x\to 0$ is the infinite
energy limit and thus can never experimentally be reached. As a
consequence of this all moments are plagued by an {\it a priori}
infinite uncertainty, which can be reduced by means of assumptions
implying that any use of the evolution equations for moments is
{\it model dependent}. The more simple solution to this problem is
to solve DGLAP equations in the $x$-space. In this framework,
besides the Laguerre method, another strategy, the so called
''brute force" method \cite{bruteforce}, represents a good
candidate. It is fundamentally a finite-differences method of
solution which reaches a good precision in the small $x$-region
\cite{Kumano,Fasching} but requires a rather large amount of
computer running time.

Here we discuss a semianalytical method, in the $x-$space, to
solve DGLAP equations \cite{noi}. It consists in constructing an
evolution operator which, depending only on the splitting
functions, can be worked out once for all. In this respect our
strategy is similar to the one in \cite{FurmPetr,CorianoSavkli}.
Our method to perform the convolutions instead, takes advantage of
an $x$ discretization (comparable to the one in
\cite{Kumano,Fasching}) which allows us to represent the evolution
operator as a matrix. Thus the procedure to construct the solution
reduces merely to a multiplication between the {\it evolution
matrix} and an {\it initial vector}, and can be done in an
extremely short computer time with the required accuracy. This is
particularly appealing in the analysis of the experimental data on
nucleon structure functions which requires a large number of
parton evolutions.\\ In the next section we discuss the (formal)
analytical solution of the DGLAP equations; in the third one the
algorithm to perform the $x$--integration is presented. The last
two sections are devoted to analyze the numerical results relative
to the evolution of polarized parton distributions, to study the
yield of our method in comparison with others and to conclude.

\section{The Evolution Operator}
\label{s:tevolution}

The DGLAP equation, up to Next-to-Leading-Order (NLO) corrections,
for the Non-Singlet distribution is\footnote{In the following we
limit ourselves to discuss the polarized parton distributions. The
application of our method to the unpolarized case is
straightforward.}
 \beqa
  \frac{\del}{\del t}
  && \Delta \tilde q_{NS}(x,t) = \nonumber\\
  &&\hspace{-8truemm}
  \left( \Delta \tilde P^{(0)}_{NS}(x) +
  \alpha(t)\Delta \tilde R_{NS}(x) \right) \otimes \Delta \tilde q_{NS}(x,t),
  \label{e:NS}
 \enqa
while for the Singlet and Gluon distributions we have:
 \beqa
 && \frac{\del}{\del t}
 \left (
 \begin{array}{c}
 \Delta \tilde q_{S}(x,t)\\
 \Delta \tilde g(x,t)
 \end{array} \right) =
 \nonumber \\
 &&\left(
 \begin{array}{cc}
 \Delta \tilde P^{(0)}_{qq}(x) & \Delta \tilde P^{(0)}_{qg}(x) \\
 \Delta \tilde P^{(0)}_{gq}(x) & \Delta \tilde P^{(0)}_{gg}(x) \\
 \end{array}
 \right)\otimes
 \left (
 \begin{array}{c}
 \Delta \tilde q_{S}(x,t)\\ \Delta \tilde g(x,t)
 \end{array} \right) + \nonumber \\
 &&\hspace{-3truemm}
 \alpha(t) \left(
 \begin{array}{cc}
 \Delta \tilde R_{qq}(x) & \Delta \tilde R_{qg}(x) \\
 \Delta \tilde R_{gq}(x) & \Delta \tilde R_{gg}(x) \\
 \end{array}
 \right) \otimes \left (
 \begin{array}{c}
 \Delta \tilde q_{S}(x,t)\\
 \Delta \tilde g(x,t)
 \end{array} \right),
 \label{e:SG}
 \enqa
where
 \bee \Delta\tilde R_{ij}(x) \equiv \Delta
 \tilde P^{(1)}_{ij}(x) - \frac{\beta_1}{2\beta_0}\Delta \tilde
 P^{(0)}_{ij}(x)\;. \label{e:deltaR}
 \ene
In these equations the symbol $\otimes$ stands for
 \bee
 f(x) \otimes g(x) \equiv \int_x^1 \frac{dy}{y}
 f\left(\frac{x}{y}\right) g(y)\;,
 \label{e:conv}
  \ene
and
 \bee
 \tilde f(x) \equiv x f(x).
 \ene
Instead of $Q^2$, we have used the variable $t$ defined by
 \bee
 t =
 -\frac{2}{\beta_0}ln\left[\frac{\alpha_s(Q^2)}{\alpha_s(Q_0^2)}\right]\;,
 \ene
where $\alpha_s$ is strong running coupling constant corrected at NLO:
 \bee
 \alpha_s(Q^2)= \frac{4 \pi}{\beta_0\,ln(Q^2/\Lambda^2)}
 \left[
 1- \frac{\beta_1\,ln(ln(Q^2/\Lambda^2))}{\beta_0^2\,ln(Q^2/\Lambda^2)}
 \right],
 \ene
and so in Eqs.~(\ref{e:NS})-(\ref{e:SG})
 \bee
 \alpha(t) \equiv
 \frac{\alpha_s(Q^2_0)}{2\pi}Exp\left\{-\frac{\beta_0}{2}t\right\}\;.
 \ene
The explicit expressions for $\beta_0$, $\beta_1$ as well as for the
Splitting Functions $\Delta P_{ij}(x)$ can be found in
\cite{SplittingFunctions}.

The equations (\ref{e:NS}) and (\ref{e:SG}) can be written in the
following general form:
 \bee \frac{\del}{\del t} {\bf f}(t) = {\bf
 \Omega}(t) \odot {\bf f}(t) \label{e:GF}
 \ene
where ${\bf f}(t)$ indicates the ``vector of components $f(x,t)$"
and ${\bf \Omega}(t)$ a linear operator acting as:
 \bee \left [
 {\bf \Omega}(t)^{~}_{~} \odot {\bf f}(t)_{}\right ]_x \equiv
 \int_x^1 ~dy~ \omega(x,y,t)\;f(y,t). \label{e:defodot}
 \ene
Note that, in the Singlet-Gluon case, ${\bf f}(t)$ becomes a doublet of vectors
and ${\bf \Omega}(t)$ a $2~$x$~2$ matrix of operators.

Due to the logarithmic dependence of $t$ on $Q^2$, the range of
values of physical interest for $t-t_0$ ($t_0$ is the starting
values of $t$, where the parton distributions are assumed known)
is small enough to expect that the Taylor's series of the solution
${\bf f}(t)$ converges rapidly. On the other hand, by deriving
repeatedly the Eq.~(\ref{e:GF}) we can write:
 \bee
 \left . \frac{\del^{k}}{\del t^k} {\bf f}(t)
 \right|_{t=t_0}= ~{\bf M}^{(k)} \odot {\bf f}(t_0),
 \ene
where the operators ${\bf M}^{(k)}$ can be obtained recursively:
 \beqa {\bf M}^{(0)} & = & {\bf I} \nonumber\\
 {\bf M}^{(1)} & = & {\bf \Omega}_0
 \nonumber\\
 {\bf M}^{(2)} & = & {\bf
 \Omega}_0^{(1)} + {\bf \Omega}_0 \odot {\bf M}^{(1)} \nonumber\\
 {\bf M}^{(3)} & = & {\bf \Omega}_0^{(2)} + 2~{\bf \Omega}_0^{(1)}
 \odot {\bf M}^{(1)} + {\bf \Omega}_0 \odot {\bf M}^{(2)}
 \nonumber\\ ... & & ....... \nonumber \\ {\bf M}^{(k)} & = &
 \sum_{i=0}^{k-1}c_i^{(k)} {\bf \Omega}_0^{(k-1-i)} \odot {\bf
 M}^{(i)}\,. \label{e:iter}
 \enqa
The $c_i^{(k)}$ indicates the $i-$th term of the $k-$th row of
Tartaglia triangle and
 \bee
 {\bf \Omega}_0 \equiv {\bf \Omega}(t_0)\;,
  \hspace{1truecm}
 {\bf \Omega}_0^{(k)} \equiv \left . \frac{\del^{k}}{\del t^k} {\bf \Omega}(t)
 \right |_{t=t_0}\;.
 \ene
Then the solution can be written as:
 \beqa
 {\bf f}(t) & = & \left ( \sum_{k=0}^{\infty} \frac{(t-t_0)^k}{ k !}
  {\bf M}^{(k)}\right ) \odot {\bf f}(t_0) \nonumber \\
  &\equiv &{\bf T}(t-t_0) \odot {\bf f}(t_0)\;,
 \label{e:solution}
 \enqa
with ${\bf T}(t-t_0)$ the Evolution Operator. As we will point out
in section {\bf\ref{s:xintegration}} the series in
Eq.~(\ref{e:solution}) converge quickly enough to obtain a very
good approximation retaining only a first few terms. It is worth
to note that if the operator ${\bf \Omega}(t)$ can be written as
$h(t)\;{\bf\Omega^{\prime}}$ (with $h(t)$ a numerical function) it
is easy to show that the series in Eq.~(\ref{e:solution}) reduces
to: \bee {\bf f}(t) = Exp\left\{\left[\int_{t_0}^t h(\tau) d\tau
\right ] {\bf\Omega^{\prime}} \right\} \odot {\bf f}(t_0)\;.
\label{e:esatta} \ene This is the case of DGLAP equation at
Leading Order (LO) approximation. Nevertheless, in
Eqs.~(\ref{e:NS})-(\ref{e:SG}), where NLO corrections are
included, we have ${\bf \Omega}(t) = {\bf \Omega}_1 + \alpha(t)
{\bf \Omega}_2\;,$ with ${\bf \Omega}_1$ and ${\bf \Omega}_2$
non-commuting operators. As a consequence the series in
Eq.~(\ref{e:solution}) cannot be summed and it is not possible
write the solution in a closed form.

\section{The $x$-Integration }
\label{s:xintegration}

The integrals in Eq.~(\ref{e:conv}) are evaluated with a method
that generalizes the one proposed in Ref. \cite{Fasching}. The
method consists to treat {\it exactly} the ``bad" behaviour of the
kernel $\omega(x,y,t)$ in Eq.~(\ref{e:defodot}) and {\it
approximate} the ``smooth" function $f(y,t)$. In particular, we
construct a $M+1$ points grid ($x_0>0,x_1,...,x_{M-1},x_M=1$) in
the interval $]0,1]$ and approximate $f(x)$ in each interval
$[x_k,x_{k+1}]$ by the cubic which fits the four point $f(x_i)$,
with $i=k-1,k,k+1,k+2$:
 \bee
 f(x) \approx \sum_{l=1}^{4}
 a_l^{(k)} x^{l-1}(x) \hspace{5truemm} \forall x \in \left[x_k,x_{k+1}\right]\;.
 \label{e:fapprox}
 \ene
The general structure of the Polarized Splitting Functions which
appear in Eqs.~(\ref{e:NS})-(\ref{e:SG}) is\footnote{The same
structure, however, holds for unpolarized and transversely
polarized splitting functions.}:
 \bee
 \Delta\tilde P(x) = \frac{{\cal A}(x)}{(1-x)_+} + {\cal B}(x)
 + \delta(1-x){\cal C}\;,
 \label{e:dp}
 \ene
and therefore the ``$i$ component" of the convolution is:
 \beqa
 && \Delta\tilde P(x_i) \otimes f(x_i)  = \nonumber \\
 && x_i\left(\int_{x_i}^1 \frac{dy}{y} \frac{ {\cal A}(x_i/y)f(y)-
 {\cal A}(1)f(x_i) }{ y - x_i } + \right.\nonumber \\
 &&\hspace{5truemm} \left. \int_{x_i}^1 \frac{dy}{y^2} {\cal
 B}\left(\frac{x_i}{y}\right)f(y) \right) + \nonumber \\
  & & \left(
 {\cal C} + {\cal A}(1)ln(1-x_i) \right ) f(x_i)\;.
 \label{e:dp-App}
 \enqa
 Substituting Eq.~(\ref{e:fapprox}) in
Eq.~(\ref{e:dp-App}) we obtain $\forall i \in \{0,...,~M-1\}$
($~\sum_{k=M}^{M-1}\equiv 0$ is understood)
 \beqa
  \Delta\tilde
  P(x_i) \otimes f(x_i) & = & \sum_{l=1}^{m}
 a_l^{(i)} \left(\beta_l^{i} + \rho_{il}^{i}\right) + \nonumber \\
 & &\sum_{k=i+1}^{M-1}\;\; \sum_{l=1}^{m} a_l^{(k)}
 \left(\gamma_{kl}^{i} + \rho_{kl}^{i} \right) + \nonumber \\
 && \hspace{-2.5truecm}
 \left(  {\cal C} + {\cal A}(1)ln(1-x_i) - {\cal A}(1) \sigma^{i}
 \right ) f(x_i)\;;
 \label{e:dpifi}
 \enqa
then we have:
 \bee
 \Delta\tilde P(x_i) \otimes f(x_i)  =
 \sum_{k=0}^{M}\omega_{ik}f(x_k),
 \label{e:final}
 \ene
where $\omega$ is the matrix of the coefficients of $f(x_k)$. The
analytical expressions for the matrices $\beta$, $\gamma$, $\rho$
and $\sigma$ can be found in \cite{noi}.

Therefore the Eqs.~(\ref{e:NS})-(\ref{e:SG}) became
\footnote{Note that $\omega^{(1)}$ matrices correspond to the
convolutions of the parton distributions with $\Delta \tilde R$ (cf
Eq.~(\ref{e:deltaR})).}:
 \beqa
 &&\hspace{-6truemm}\frac{\del}{\del t} \Delta \tilde q_{NS}(x_i,t) = \nonumber\\
 &&\hspace{-6truemm}\sum_{k=0}^M\left( \omega^{(0)~NS}_{ik} +
 \alpha(t)\omega^{(1)~NS}_{ik} \right) \Delta \tilde q_{NS}(x_k,t)
 \label{e:NS-om}
 \enqa
 \beqa
 &&\frac{\del}{\del t} \left (
 \begin{array}{c}
 \Delta \tilde q_{S}(x_i,t)\\
 \Delta \tilde g(x_i,t)
 \end{array} \right) =
 \sum_{k=0}^{M}\left [ \left(
 \begin{array}{cc}
 \omega^{(0)~qq}_{ik} & \omega^{(0)~qg}_{ik}\\
 \omega^{(0)~gq}_{ik} & \omega^{(0)~gg}_{ik} \\
 \end{array}
 \right)\right.\nonumber \\
 &&\hspace{-.9truemm}\left. + \alpha(t)
 \left(
 \begin{array}{cc}
 \omega^{(1)~qq}_{ik} & \omega^{(1)~qg}_{ik} \\
 \omega^{(1)~gq}_{ik} & \omega^{(1)~gg}_{ik} \\
 \end{array}
 \right)
 \right]
 \left (
 \begin{array}{c}
 \Delta \tilde q_{S}(x_k,t)\\
 \Delta \tilde g(x_k,t)
 \end{array} \right).
 \label{e:SG-om}
 \enqa
We solve these equations by means of the method shown in section
{\bf\ref{s:tevolution}}: the operator $\Omega(t)$ and then the ${\bf
M}^{(k)}$ became now numerical matrices, and the symbol $\odot$ stands
for the usual rows by columns product. We would stress the fact that the
matrices ${\bf M}^{(k)}$ depend only on the points $x_i$ and so they can
be numerically evaluated once for all.

\section{Numerical Analysis}
\label{s:nunan}

The convergence of our algorithm is controlled by two parameters: the
order $n$ of the truncated series
\bee
{\bf T}^{(n)}(t-t_0) =
\sum_{k=0}^{n} \frac{(t-t_0)^k}{ k !}  {\bf M}^{(k)}\;,
\ene
which define the evolution operator, and the number $M$ of the points of
$x-$integration.

\begin{figure}[t]
 \epsfig{file=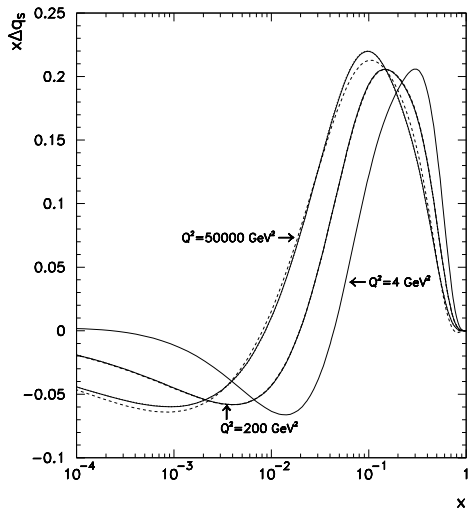,height=7.4cm}
 \caption{ The initial Singlet
 distribution ($Q^2=4~GeV^2$, solid line) and the evolved ones for
 $n=3$ (dashed lines), $n=6$ (dotted lines) and $n=12$ (solid
 lines) corresponding at $Q^2=200~GeV^2$ and $Q^2=50000~GeV^2$. We
 use $M=100$. }
 \label{f:fig1}
 \end{figure}

To test the accuracy of our method we evolve the Gehrmann and
Stirling polarized singlet-gluon initial distributions (cf
\cite{GS}) from $Q_0^2 = 4~GeV^2$ $(t_0=0)$ to $Q^2
=200~GeV^2~(t=0.136)$ and $Q^2=50000~GeV^2~(t=0.245)$. We choose
to work, as in the paper \cite{Kumano}, in the fixed flavour
scheme, $n_f=3$, with $\Lambda^{(4)}_{QCD}=231~MeV$, and without
taking into account, in the $Q^2$ evolution of $\alpha_s$, quark
thresholds. The range $]0,1]$ has been divided in $M$ steps by
$M+1$ points: $x_0, x_1, ..., x_M$ distributed in such a way that
the function $ln(x) + 2x$ varies by the same amount at any step;
this function is slightly different from the pure logarithmic
distribution commonly used in literature \cite{Kumano,Fasching},
but allow, in our case, a more uniform distribution of the
numerical errors. The end points are fixed to be $x_0=1\times
10^{-8}$ and $x_M=1$; however, for a better reading, in the
Figures~{\bf\ref{f:fig1}}$-${\bf\ref{f:fig4}} the $x-$axis ranges
from $10^{-4}$ to $1$.

\begin{figure}[t]
 \epsfig{file=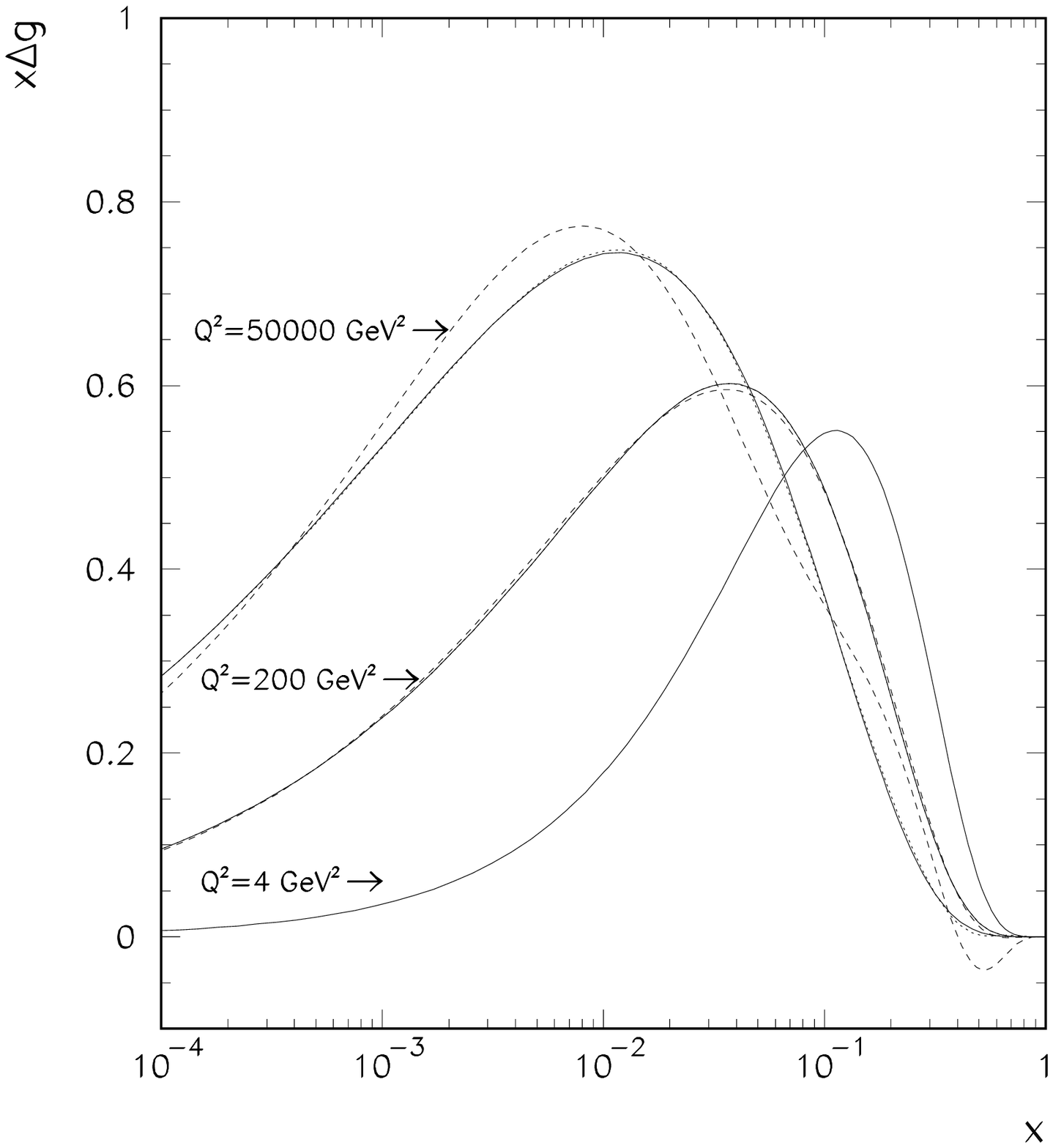,height=7.4cm}
 \caption{The same in Fig. \protect\ref{f:fig1} for Gluons.}
 \label{f:fig2}
 \end{figure}

First, we fix $M= 100$. In Figs.
{\bf\ref{f:fig1}}$-${\bf\ref{f:fig2}} are reported
the evolved singlet and gluon distributions, respectively, obtained with
$n=3,~6~$and$~12$ for $Q^2 =200~GeV^2$ and $Q^2 =50000~GeV^2$. It is worth
to note the very fast convergence of the series to the solution, as
already observed above. As a matter of fact, the maximum difference
between the solutions relative to $n=6$ and $n=12$ is $1.4\times
10^{-5}~$($7.6\times 10^{-4}$) for the singlet, and $9.3\times
10^{-5}~$($5.3\times 10^{-3}$) for the gluon distribution, in
correspondence of $Q^2=200~GeV^2~$($Q^2=50000~GeV^2$).

\begin{figure}[t]
 \epsfig{file=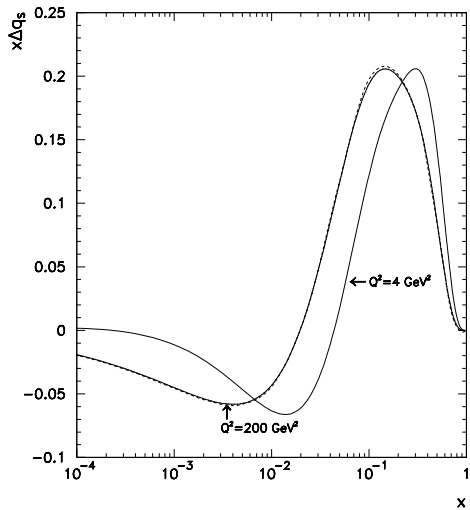,height=7.4cm}
 \caption{ The initial Singlet
 distribution ($Q^2=4~GeV^2$, solid line) and the evolved one at
 $Q^2=200~GeV^2$ with $M=100$ (solid line), $M=50$ (dotted line)
 and $M=25$ (dashed line) with $n=12$. }
 \label{f:fig3}
 \end{figure}

\begin{figure}[t]
 \epsfig{file=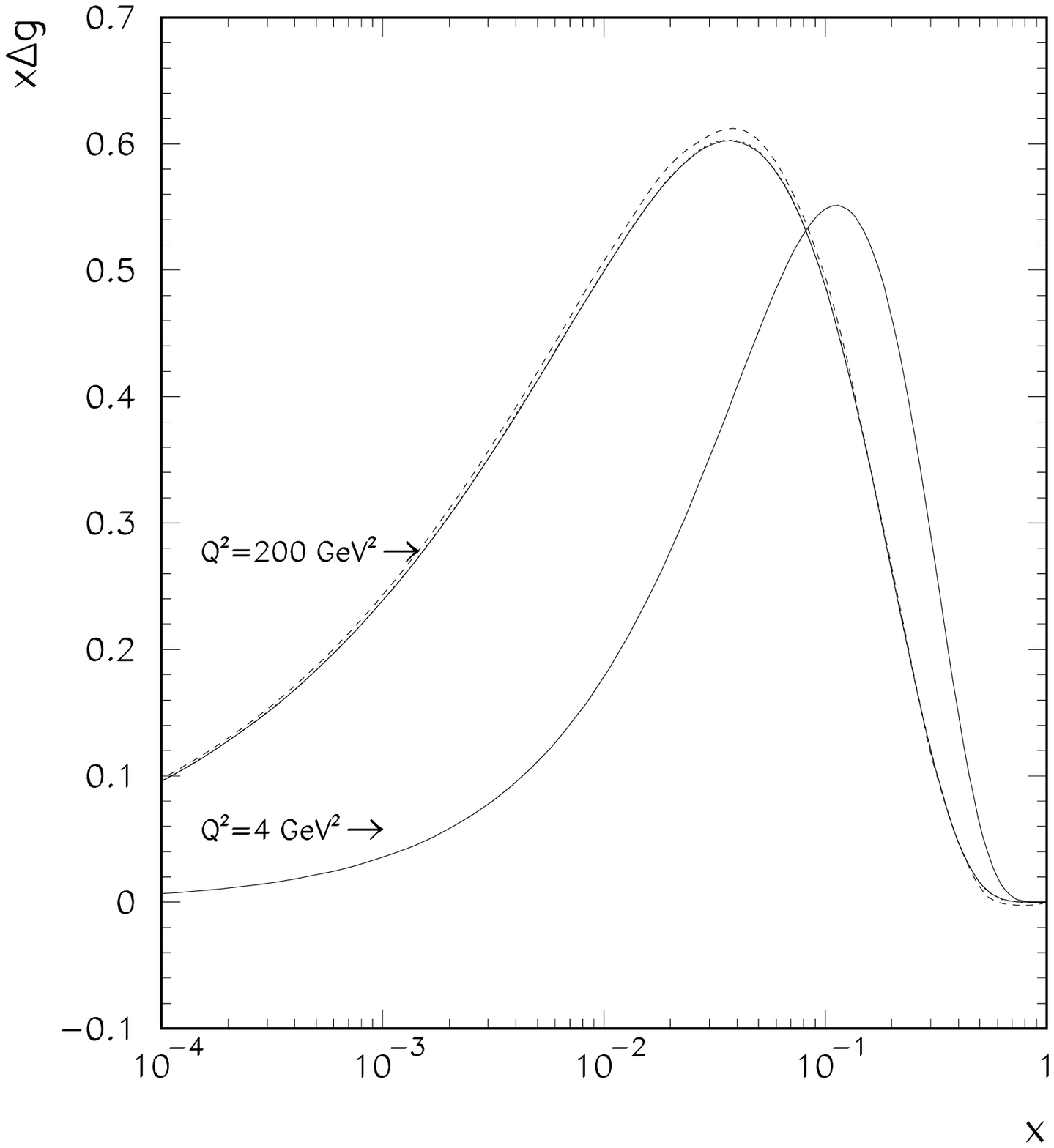,height=7.4cm}
 \caption{ The same in Fig.
 \protect\ref{f:fig3} for Gluons. }
 \label{f:fig4}
 \end{figure}

Next we fix $n=12$ and $Q^2= 200~GeV^2$. In Figs.
{\bf\ref{f:fig3}}$-${\bf\ref{f:fig4}} are plotted the approximated
evolved distributions with $M=25,~50,~100$: the maximum difference
on the common points between $M=50$ and $M=100$ is $4.6 \times
10^{-4}$ for the singlet and $7.9 \times 10^{-4}$ for the gluons.
By comparing the results in Fig.
{\bf\ref{f:fig3}}$-${\bf\ref{f:fig4}} with the corresponding Figs.
{\bf 1}-{\bf 4} in Ref. \cite{Kumano}, we observe, besides a good
numerical agreement of the results, a faster convergence as the
number M of integration points increases, as a consequence of our
more accurate $x$-integration procedure with respect to the so
called ``brute force'' methods. In fact it should be observed that
reducing from the cubic to the linear approximation of $f(x)$ in
Eq.~(\ref{e:fapprox}), the accuracy ${\cal E}(x,t)$ (defined in
the sequel) becomes about 1 and 3 order of magnitude bigger,
respectively for singlet and gluon.

\begin{figure}[t]
 \epsfig{file=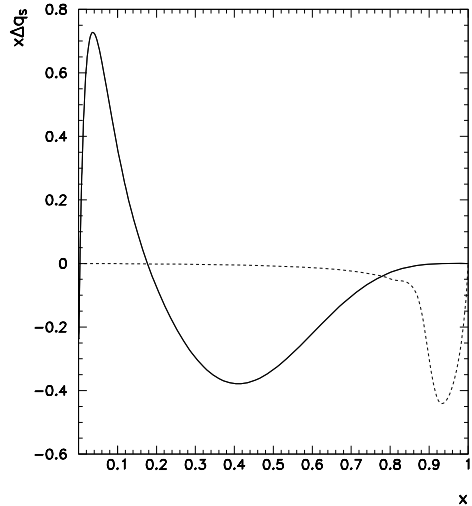,height=7.4cm}
 \caption{ For $n=12$, $M=100$
 and $Q^2 = 200~GeV^2$ both sides of the Eq.~(\protect\ref{e:SG})
 are plotted (solid line and dotted line), in correspondence of the
 Singlet distribution. Dashed line represents ${\cal
 E}(x)~\times~10^{3}$ (see text).}
 \label{f:fig5}
 \end{figure}

\begin{figure}[t]
 \epsfig{file=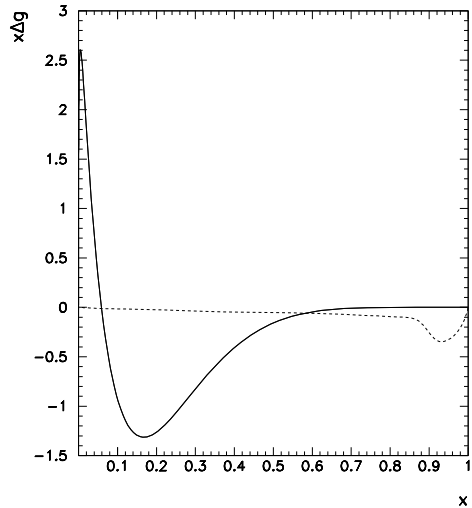,height=7.4cm}
 \caption{The same in Fig. \protect\ref{f:fig5} for Gluons.
 Dashed line represents ${\cal E}(x)~\times~10^{4}$ (see text).}
 \label{f:fig6}
 \end{figure}

To discuss the degree of accuracy of the method in \cite{noi} were
introduced a global accuracy ${\cal E}(x,t)$ defined as the
difference between left and right-hand side of the
Eq.~(\ref{e:SG}). The comparison between the range of values of
${\cal E}$ with the one of both sides of Eq.~(\ref{e:SG})
represents a very good estimate of the degree of accuracy of the
solution. In Figs. {\bf\ref{f:fig5}} and {\bf\ref{f:fig6}} are
plotted, for $n=12$, $M=100$ and $Q^2 = 200~GeV^2$ both sides of
the Eq.~(\ref{e:SG}) and the corresponding (rescaled) accuracy
${\cal E}(x,t)$ \footnote{The integration in the right-hand side
has been performed numerically after an $x$-interpolation of the
discrete values obtained with the evolution operator, while the
left-hand side is worked out by direct derivation of
Eq.~(\ref{e:solution}).}. It appears evident that an excellent
approximation of the solution is obtained.

Another advantage of our method, once fixed the accuracy of the
solution, appears to be the running time to get each evolution. In
fact, the simple analytical structure of the evolution matrix
${\bf T}$ makes the solution procedure considerably fast. As a
matter of fact, once given the Splitting Functions and constructed
the corresponding matrices ${\bf M}^{(k)}$ (we have used
Mathematica \cite{Mathematica} to do this), a single evolution,
i.e. the multiplication of the ${\bf T}$ evolution matrix by the
initial vector, require, for $n=12$ and $M=100$, about $6 \times
10^{-2}~sec$ on an AlphaServer 1000 using a Fortran Code.

Particularly interesting is the comparison between our method and the
one presented in \cite{FurmPetr,CorianoSavkli}, where an evolution
operator is also introduced. Firstly we observe that the latter method
is based on a polynomial expansion of the splitting and distribution
functions. The expansion is equivalent to an expansion in power of $x$.
As a consequence it is affected by problems of convergence for
$x\rightarrow 0$, due to the branch point in zero of the involved
functions. This is the source of the difficult encountered in the
small-$x$ region, which are not present in our approach in which an
optimized Newton-Cotes-like quadrature formula is employed.\\
Second, also in the $x$-region of convergence, the Laguerre polynomial
expansion need, for each evolution process, the computation of the
moments of the initial parton distributions with respect to the
polynomials: this procedure requires a remarkable amount of CPU-time
with respect to our approach in which only the evaluation of the initial
parton distribution in the $M$ grid points is needed.

\section{Conclusions}
\label{s:conclu}

We have discussed a new algorithm to solve the DGLAP evolution
equations, in the $x$-space, which appears suitable for a rather
large class of coupled integrodifferential equations.

The method produces a solution which is analytical in the
$Q^2$-evolution parameter and approximate, but rapidly convergent,
in the $x-$space. It allows to construct, once for all, an
evolution operator in matrix form. It depends only on the
splitting functions appearing in the equations and can be rapidly
applied to whatever initial distribution to furnish the evolved
one, requiring for each evolution only a few hundredth of second.

It is worth to note the reliability of our $x$--integration
algorithm, which gets excellent approximations on the whole
$x-$range (we use, for all the calculations, $10^{-8}\le$ $x$
$\le$ 1), also with few integration points, resulting in an
evolution matrix of particularly small dimensions.

In conclusion, our method, whose numerical implementation is
straightforward, appears to be very fast, very accurate and extremely
stable with respect to the increasing of convergence parameters (i.e.
$n$, the order of the truncated series which gives the Evolution
Operator, and $M$, the number of integration points). For these reasons
it represents a powerful tool to analyze the experimental data on
nucleon structure functions.

\vspace{1truecm}

 \acknowledgments
One of us (P.S.) thanks G. Altarelli for useful discussions during
the Corf\'u Conference.


%

\end{document}